\documentstyle[prl,aps,epsfig,twocolumn]{revtex}

\begin{document}
\draft

\twocolumn[\hsize\textwidth\columnwidth\hsize\csname@twocolumnfalse\endcsname
\title{Induced Magnetic Monopole from Trapped $\Lambda$-Type Atom }
\author{P. Zhang, Y. Li and C. P. Sun$^{a,b}$}
\address{Institute of Theoretical Physics, the Chinese Academy of Sciences, Beijing, 100080, China}
 \maketitle

\begin{abstract}

We investigate the spatial motion of the trapped atom with the
electromagnetically induced transparency (EIT) configuration where
the two Rabi transitions are coupled to two classical light fields
respectively with the same detuning. When the internal degrees of
freedom can be decoupled adiabatically from the spatial motion of
the center of mass via the Born-Oppenheimer approximation, it is
demonstrated that the lights of certain profile can provide the
atom with an effective field of magnetic monopole, which is the
so-called induced gauge field relevant to the Berry's phase. Such
an artificial magnetic monopole structure manifests itself in the
characterizing energy spectrum.
\end{abstract}
  ]

Modern concept of magnetic monopole in quantum mechanics was
postulated by Dirac in 1931 \cite{dirac}. Since that time physicist
have been making efforts to seek for the magnetic monopole in real
space for more than seventy years. Though convincing evidence for
its existence in real space has not yet been found, the theoretical
conception of magnetic monopole has initiated many important
progresses in both physics and mathematics. In fact, in the
extremely high energy scale that we have not reached at present,
the grand unification theory \cite{gut} predicted the magnetic
monopole as a consequence of the beautiful topology structure of
Yang-Mills theory \cite{yang1}. The discovery of Berry phase also
resulted in a
physical implementation  of magnetic monopole in the parameter space \cite%
{BPF}. Precisely speaking, the degeneracy point in the parameter
space acts like a magnetic monopole caused by an effective gauge
field. The Berry phase based magnetic monopole of this kind can be
demonstrated in association with the anomalous Hall effect of
ferromagnetic metals \cite{FZ} where the slowly changing parameter
is just the crystal momentum. Now we consider an artificial
realization of magnetic monopole in real space.

For the quantum adiabatic process induced Berry phase, it is well
known that when the slowly varying parameters are the dynamic
variables of a subsystem interacting with another subsystem with
fast varying variables, the adiabatic separation of the two
subsystems via the Born-Oppenheimer approximation \cite{boa} can
provide the slow subspace with a scaler and a vector potentials
\cite{WZK} called the induced gauge potential. In the neutron spin
precession experiment \cite{prd}the Aharonov-Borhm effect caused by
this vector potential was pointed out as a manifestation of Berry
phase. In this article, we will derive the monopole type induced
gauge field for the spatial motion of $\Lambda -$type atom
interacting with control and
probe laser beams which drive the transitions $|e\rangle $-$|1\rangle $ and$%
|e\rangle $-$|2\rangle $ respectively (see Fig. 1)
\cite{EIT,Sun-prl}. We will also show that when the atom is cold
enough, its internal degrees of freedom can decouple adiabatically
from the spatial motion of the center of mass. Correspondingly, the
Born-Oppenheimer approximation provides the atomic center of mass
with an effective magnetic monopole field as a special induced
gauge field relevant to the Berry phase when the lights are
artificially shaped in certain profiles. We predict that such an
artificial magnetic monopole can be observed experimentally through
its special spectral structure. In fact, the prompt advance in
experiments of trapping and cooling atoms has provided a platform
to test our predictions exactly.

The Hamiltonian for our cold atom system driven by two laser beams
(see Fig. 1) can be written in the form $H={\bf
P}^{2}/2M+H_{f}({\bf r})$ with the local internal Hamiltonian
\begin{eqnarray}
H_{f}({\bf r}) &=&\sum_{i=1}^{2}V_{i}\left( {\bf r}\right) \left\vert
i\right\rangle \left\langle i\right\vert +\Delta \left\vert e\right\rangle
\left\langle e\right\vert  \nonumber \\
&&+\Omega _{p}\left( {\bf r}\right) \left\vert e\right\rangle \left\langle
1\right\vert +\Omega _{c}\left( {\bf r}\right) \left\vert e\right\rangle
\left\langle 2\right\vert +h.c..
\end{eqnarray}%
Here, ${\bf r}$ is the atomic position, $\Delta $ the one-photon
detunning, $\Omega _{c}\left( {\bf r}\right) $ ($\Omega _{p}\left(
{\bf r}\right)$) the Rabi frequency of the probe (control) beam,
and $V_{i}\left( {\bf r}\right) $ the trap potential of the $i$-th
inner energy level. We assume that there is no trap for the level $e$, i.e., $%
V_{e}\left( {\bf r}\right) =0$, and there is the same trap for both
level $1$ and $2$, i.e., $V_{1}\left({\bf r}\right) =V_{2}\left(
{\bf r}\right) =V\left( {\bf r}\right)$. This assumption just
ensures the occupance of the dark state.

\begin{figure}[h]
\hspace{24pt}\includegraphics[width=6cm,height=4cm]{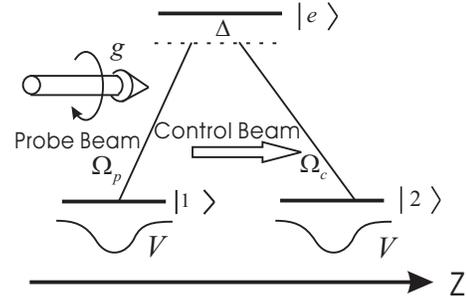}
\caption{Three level atoms interacting with two laser beams. The
probe beam coupling the states $|e\rangle $ and $|1\rangle $ has an
orbital angular momentum. The atoms in state $|1\rangle $ and
$|2\rangle $ are trapped by the potential $V$, which is a function
of the atomic position ${\bf r}$. }
\end{figure}

To use the generalized Born-Oppenheimer approximation\cite{prd}, we
first diagonalize the interaction part $H_{f}({\bf r})$ of the
Hamiltonian and obtain the ${\bf r}$-dependent eigenvalues of
$H_{f}({\bf r})$: $E_{0}\left( {\bf r}\right) =V\left( {\bf
r}\right)$ and
\begin{equation}
E_{\pm }\left( {\bf r}\right) =\frac{1}{2}(\tilde{\Delta}\pm \sqrt{%
4\left\vert \Omega _{c}\right\vert ^{2}+4\left\vert \Omega _{p}\right\vert
^{2}+\tilde{\Delta}^{2}})+V\left( {\bf r}\right) .
\end{equation}%
where $\tilde{\Delta}=\tilde{\Delta}\left( {\bf r}\right) =\Delta
-V\left( {\bf r}\right) $ is the local one-photon detunning. The
eigenstate corresponding to $E_{0}\left( {\bf r}\right) $ is the
${\bf r}$-dependent dark sate defined as
\begin{equation}
\left\vert D\left( {\bf r}\right) \right\rangle =\frac{1}{\Omega }\left[
\Omega _{p}\left\vert 2\right\rangle -\Omega _{c}\left\vert 1\right\rangle %
\right] .
\end{equation}%
where $\Omega =\Omega \left( {\bf r}\right) =\sqrt{\left\vert
\Omega _{c}\right\vert ^{2}+\left\vert \Omega _{p}\right\vert
^{2}}$. The other two eigenstates corresponding to the
eigenenergies $E_{\pm }\left( {\bf r}\right) $ can be noted as
$\left\vert B_{\pm }\left( {\bf r}\right) \right\rangle $. Their
explicit expressions are not necessary to the following discussion
for the case that the atom spatial motion is sufficiently slow so
that the internal motion is not excited.

It is well known that the atomic wave function in the ${\bf r}$%
-representation can be written as%
\begin{equation}
\left\langle {\bf r}\right\vert \Psi \rangle =\psi _{0}\left( {\bf r}\right)
\left\vert D\left( {\bf r}\right) \right\rangle +\sum_{k=+,-}\psi _{k}\left(
{\bf r}\right) \left\vert B_{k}\left( {\bf r}\right) \right\rangle .
\end{equation}%
When the energy gaps $E_{\pm }-E_{0}\left( {\bf r}\right) $ between the dark
state $\left\vert D\left( {\bf r}\right) \right\rangle $ and the states $%
\left\vert B_{\pm }\left( {\bf r}\right) \right\rangle $ are large
enough, the Born-Oppenheimer approximation is applicable. Under
this approximation, the atom can be assumed to be "kept" in the
dark state $\left\vert D\left( {\bf r}\right) \right\rangle $ at
every position ${\bf r}$ and the eigen wave function of the total
Hamiltonian $H$ can be written as $\left\langle {\bf r}\right\vert
\Psi \rangle =\psi _{0}\left( {\bf r}\right) \left\vert D\left(
{\bf r}\right) \right\rangle $ where $\psi _{0}\left( {\bf
r}\right)
$ satisfies the eigenequation%
\begin{equation}
\left[ \frac{1}{2M}\left( {\bf P}+{\bf A}_{0}\right) ^{2}+V\left( {\bf r}%
\right) \right] \psi _{0}\left( {\bf r}\right) =E\psi _{0}\left( {\bf r}%
\right) .  \label{a}
\end{equation}%
Here, ${\bf A}_{0}\left( {\bf r}\right) =-i\left\langle D\left( {\bf r}%
\right) \right\vert \nabla \left\vert D\left( {\bf r}\right) \right\rangle $
is the induced gauge potential corresponding to the dark state.

As in Ref. \cite{fermion}, we may assume that the probe beam and
the control beam have the same frequency and propagate along the
$z$ direction. Then the Rabi frequencies $\Omega _{p}\left( {\bf
r}\right) $ and $\Omega _{c}\left( {\bf r}\right) $ can be
expressed as
\begin{eqnarray}
\Omega _{p} &=&\left\vert \Omega _{p}\right\vert \exp \left( i{\bf k}\cdot
{\bf r}+g\phi \right) ,  \label{b} \\
\Omega _{c} &=&\left\vert \Omega _{c}\right\vert \exp \left( i{\bf k}\cdot
{\bf r}\right) .  \nonumber
\end{eqnarray}%
Here, the real parameters $\left\vert \Omega _{p}\left( {\bf r}\right)
\right\vert $ and $\left\vert \Omega _{c}\left( {\bf r}\right) \right\vert $
are just the slowly varying norms of the $\Omega _{p}$ and $\Omega _{c}$ and
$\phi $ the directional angle of the $x-y$ plane. Note that in writing down
Eq. (\ref{b}), we have also assumed that the probe beam has an orbital
angular momentum $g\phi $ \ with $g$ an integer \cite{al}. Then the induced
gauge potential ${\bf A}_{0}$ of the dark state can be written as%
\begin{equation}
{\bf A}_{0}=\nabla \phi _{c}+\frac{g\left\vert \Omega _{p}\right\vert ^{2}}{%
\Omega ^{2}}\nabla \phi .
\end{equation}%
It is obvious that the orbital angular momentum number $g$ of the probe beam
induces an effective magnetic monopole of strength $g.$ We thus conclude
that the effective magnetic charge of this induced monopole can be
controlled artificially by adjusting the angular momentum of photons in the
probe beam.

When the norms $\left\vert \Omega _{p}\left( {\bf r}\right)
\right\vert $ and $\left\vert \Omega _{c}\left( {\bf r}\right)
\right\vert $ of the Rabi frequencies take the forms $\left\vert
\Omega _{p}\right\vert ^{2}=\xi \left( r+z\right) $, $\left\vert
\Omega _{c}\right\vert ^{2}=\xi \left( r-z\right) $, the potential
${\bf A}_{0}$ has the same form as that of the potential created by
a monopole and can be expressed as
\begin{equation}
{\bf A}_{0}=\nabla \phi _{c}+\frac{g\left( 1+\cos \theta \right) }{2r\sin
\theta }{\bf e}_{\phi }.
\end{equation}%
Here, $r$, $\theta $ and $\phi $ are just the spherical polar
coordinates. In this case we have $\left\vert \Omega
_{p}\right\vert =\left\vert \Omega _{c}\right\vert =0$ at the
origin ${\bf r}=0$. This leads to an "accidental degeneracy"
$E_{\pm }\left( {\bf 0}\right) =E_{0}\left( {\bf 0}\right) $ at the
origin where the energy levels cross. Thus the Born-Oppenheimer
approximation does not work well. For this reason, in the following
discussion, we only discuss the atomic motion around the region far
away from the origin.

\begin{figure}[h]
\hspace{54pt}
\includegraphics[width=3.5cm,height=3.5cm]{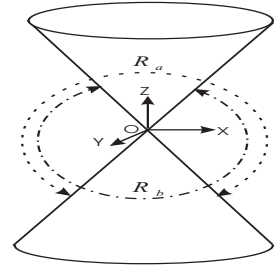}
\vspace{0.5cm}
\caption{The whole space excluding the origin $O$ is divided into two regions $%
R_{a}$ and $R_{b}$. $R_{a}$ is the space excluding the lower
circular cone (the region enveloped by the dashed line), $R_{b}$
is the space excluding the upper one (the region enveloped by the
dash dot line). The space outside of the two cones is the overlap
of $R_{a}$ and $R_{b}$.}
\end{figure}

For the original Dirac's monopole, as is well known, the induced
monopole potential ${\bf A}_{0}$ has singularity at the string of
$\theta =0$. But this singularity can be exorcized by means of the
approach developed by Wu and Yang \cite{yang1,yang2}. To this end
one needs to divide the total real
space (except the origin) into two regions (see Fig. 2), $R_{a}: 0\leq \theta <%
\frac{\pi }{2}+\delta $ and $R_{b}: \frac{\pi }{2}-\delta \leq
\theta <\pi $, which are two overlapping caps and have their well
defined local coordinates. Then the dark state $\left\vert D\left(
{\bf r}\right) \right\rangle $ and the wave function $\psi
_{0}\left( {\bf r}\right) $ can no longer be defined globally.
Instead, $\left\vert D\left( {\bf r}\right) \right\rangle $ and
$\psi _{0}\left( {\bf r}\right) $ will have different expressions
in the different regions marked by $a$ and $b$. For instance the
dark state can be defined as%
\begin{equation}
\left\vert D\left( x\right) \right\rangle _{a}=e^{-i\left( \phi
_{c}+g\phi \right) }\frac{1}{\Omega }(\Omega _{p}\left\vert
2\right\rangle -\Omega _{c}\left\vert 1\right\rangle )
\end{equation}%
in $R_{a}$ and \qquad \qquad
\begin{equation}
\left\vert D\left( x\right) \right\rangle _{b}=e^{-i\phi _{c}}\frac{1}{%
\Omega }(\Omega _{p}\left\vert 2\right\rangle -\Omega
_{c}\left\vert 1\right\rangle )
\end{equation}%
in $R_{b}$. The Schroedinger equation (\ref{a}) can be rewritten in
different caps as
\begin{equation}
\left[ \frac{1}{2M}\left( {\bf P}+{\bf A}_{0\alpha }\right)
^{2}+V\left( {\bf r}\right) \right] \psi _{0\alpha }\left( {\bf
r}\right) =E\psi _{0\alpha }\left( {\bf r}\right)   \label{ss}
\end{equation}%
in $R_{\alpha }$ for $\alpha =a,b$. Here, $\psi _{0a}\left( {\bf
r}\right) $ and $\psi
_{0b}\left( {\bf r}\right) $ are the expressions of $\psi _{0}\left( {\bf r}%
\right) $ in $R_{a}$ and $R_{b}$ respectively and the gauge potential ${\bf A%
}_{0a}$ (${\bf A}_{0b}$) can be expressed as%
\begin{equation}
{\bf A}_{0a}=\frac{g\left( -1+\cos \theta \right) }{2r\sin \theta }\hat{e}%
_{\phi },\ \ {\bf A}_{0b}=\frac{g\left( 1+\cos \theta \right) }{2r\sin \theta }%
\hat{e}_{\phi }.
\end{equation}%
Apparently, ${\bf A}_{0a}$ (${\bf A}_{0b}$) is not singular in the region $%
R_{a}$ ($R_{b}$).  In the overlap of $R_{a}$ and $R_{b}$, we have
a connection $\psi _{0b}\left( {\bf r}\right) =\psi _{0a}\left(
{\bf r}\right) e^{-ig\phi }$ due to the $U(1)-guage.$ Such two
local wave functions in two caps with this connection  in the
overlapping region are mathematically called the wave section.

We now consider a simple case that the trap potential $V\left( {\bf r}%
\right) $ is spherically symmetrical. In this case we can separate
the radical degree of freedom $r$ and the angular degrees of
freedom $\theta $ and $\phi $  in the Schroedinger equation
(\ref{ss}) by invoking the
generalized angular momentum operator \cite{yang2}%
\begin{equation}
{\bf L}={\bf r}\times \left( {\bf P}+{\bf A}_{0}\right)
-\frac{g{\bf r}}{2r}.
\end{equation}%
It was proved \cite{yang2} that $\psi _{0}\left( {\bf r}\right) =R_{l}\left(
r\right) Y_{q,l,m}$ can be expressed in terms of the monopole harmonics $Y_{%
\frac{g}{2},l,m}\left( \theta ,\phi \right) $ which is the common "{\it %
eigensection}" of $L^{2}$ and $L_{z}$ with respect to the eigenvalues $%
l\left( l+1\right) $ and $m$. Here, we have $l=\left\vert \frac{g}{2}%
\right\vert ,\left\vert \frac{g}{2}\right\vert +1,...$ and
$m=-l,-l+1,...,l$.
The radical wave function $R_{l}\left( r\right) $ satisfies the equation%
\begin{equation}
\left[ -\frac{\partial _{r}\left( r^{2}\partial _{r}\right)}{2Mr^{2}} +%
\frac{l\left( l+1\right) -\left( g/2\right) ^{2}}{2Mr^{2}}+V\left(
r\right) -E\right] R_{l}=0.  \label{d}
\end{equation}%
In the special case $V=0$, when $E>0$, the solution of Eq. (\ref{d}) is \cite%
{v=0} a Bessel function $R=\frac{1}{\sqrt{kr}}J_{\mu }\left( kr\right) $ with%
\begin{equation}
\mu =\sqrt{l\left( l+1\right) -\left( \frac{g}{2}\right) ^{2}+\frac{1}{4}},\ \ k=%
\sqrt{2ME}.
\end{equation}

Next we consider a more interesting case. We assume the the trap is a
harmonic potential%
\begin{equation}
V=\frac{1}{2}M\omega _{z}^{2}\left( z-z_{0}\right) ^{2}+\frac{1}{2}M\omega
^{2}\rho ^{2}
\end{equation}%
where $z_{0}>0$ and $\rho ^{2}=x^{2}+y^{2}$ (see Fig. 3). In this
case, the atom is confined near a fixed point $\left(
0,0,z_{0}\right) $ in the region $R_{a}$. Therefore, we need only
consider the expression $\psi _{0a}$ of $\psi _{0}$ in $R_{a}$.
Since the trap potential is cylindrically symmetrical, it is
convenient to discuss this problem with the cylindrical coordinate
$\left( \rho ,z,\phi \right) $. Because of the cylindrical symmetry
of $V$ and ${\bf A}_{0}$, the wave function $\psi _{0a}$ has the
factor $\exp \left( im\phi \right) $ $\left( m=0,\pm 1,\pm
2,...\right) $
which is just the generator of the rotation along $z$ axis. Then we have $%
\psi _{0a}=T_{m}\left( \rho ,z\right) \exp \left( im\phi \right) $ where $%
R_{m}\left( \rho ,z\right) $ satisfies the radical Schroedinger equation%
\begin{equation}
\frac{1}{2M}\left[ -\partial _{\rho }^{2}-\frac{1}{\rho }\partial _{\rho
}-\partial _{z}^{2}+F_{m}\left( \rho ,z\right) \right] T_{m}+VT_{m}=ET_{m}.
\label{e}
\end{equation}%
Here, the function $F_{m}\left( \rho ,z\right) $ is defined as%
\begin{equation}
F_{m}\left( \rho ,z\right) =\left( \frac{m}{\rho }+g\frac{z-\sqrt{\rho
^{2}+z^{2}}}{2\sqrt{\rho ^{2}+z^{2}}\rho }\right) ^{2}.
\end{equation}

\begin{figure}[h]
\hspace{2cm}
\includegraphics[width=4cm,height=3.5cm]{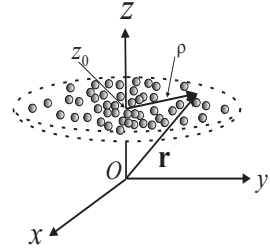}
\vspace{0.5cm}

\caption{The atoms are trapped in the "discus form" region whose
center is the point $(0,0,z_0)$. In this figure, $O$ is the origin
of the coordinate.}
\end{figure}

If the trap along $z$ axis is strong enough, i.e., $\omega _{z}$ is
large enough, we can make the approximation $F_{m}\left( \rho
,z\right) \approx F_{m}\left( \rho ,z_{0}\right) $. It is obvious
that $F_{m}\left( \rho ,z_{0}\right) $ can be expanded as a Laurent
series of $\rho $. We also assume that the trap localized in $x-y$
plane is also strong enough that we can only keep $F_{m}\left( \rho
,z_{0}\right) $ up to the term proportional
to $\rho ^{2}$. Thus approximately we have%
\begin{equation}
F_{m}\left( \rho ,z\right) \approx \frac{m^{2}}{\rho ^{2}}+\frac{g^{2}}{%
16z_{0}^{4}}\rho ^{2}-\frac{mg}{2z_{0}^{2}},
\end{equation}%
and we can solve the Eq. (\ref{e}) to obtain the energy spectrum
\begin{equation}
E_{m,n_{\rho },n_{z}}=\left( 2n_{\rho }+\left\vert m\right\vert +1\right)
\widetilde{\omega }-\frac{mg}{4Mz_{0}^{2}}+\left( n_{z}+\frac{1}{2}\right)
\omega _{z}
\end{equation}%
in terms of the radical quantum number $n_{\rho }$ and the vertical one $%
n_{z}$ ($=0,1,2,...)$, where
\begin{equation}
\widetilde{\omega }=\sqrt{\omega ^{2}+\frac{g^{2}}{16M^{2}z_{0}^{4}}}
\end{equation}%
is the modified radical frequency for the two dimensional reduced radical
oscillator. The corresponding radical frequency shift $\widetilde{\omega }%
-\omega $ $\sim $ $-\frac{mg}{4Mz_{0}^{2}}$ can be regarded as the first
observable effect of the artificial magnetic monopole. The additional term $-%
\frac{mg}{4Mz_{0}^{2}}$ in the energy spectrum of the spatial
motion of atom may reflect its effect in realistic experiment. The
corresponding wave
function can also be obtained explicitly:%
\begin{eqnarray}
\psi _{0a}^{m,n_{\rho },n_{z}} &=&N_{n_{z}}e^{im\phi }\rho ^{\left\vert
m\right\vert }e^{-M\left( \widetilde{\omega }+\omega _{z}\right) \rho ^{2}/2}
\\
&&\times F\left( -n_{\rho },\left\vert m\right\vert +1,M\widetilde{\omega }%
\rho ^{2}\right) H_{n_{z}}\left( \sqrt{M\omega _{z}}z\right) .  \nonumber
\end{eqnarray}%
Here, $N_{n_{z}}=\left[ \sqrt{M\omega _{z}}/\sqrt{\pi }2^{n_{z}}n_{z}!\right]
^{\frac{1}{2}}$, $F$ is the confluent hypergeometric function and $H_{n_{z}}$
the Hermit function.

Since the above results are achieved with the generalized
Born-Oppenheimer approximation , we should investigate the
condition under which this approximation is applicable. The
adiabatic condition can be obtained semiclassically. For
simplicity, we only consider the case $\Delta =0$. In our problem,
the sufficient condition of adiabatic approximation is
\begin{equation}
\left\vert \left\langle D\right\vert \nabla \left\vert B_{\pm
}\right\rangle \cdot {\bf v}|\cdot|E_{\pm }-E_{0}\right\vert
^{-1}<<1, \label{ab}
\end{equation}%
where ${\bf v}$ is the velocity of the atomic center of mass (c.m).
By straightforward calculation, it can be obtained that the upper
limit of $\left\vert \left\langle D\right\vert \nabla \left\vert
B_{\pm }\right\rangle \cdot {\bf v}\right\vert $ is
$\frac{1}{2r}(\left\vert v_{\rho }\right\vert +\left\vert
v_{z}\right\vert +g\left\vert v_{\phi }\right\vert )\sim gv/r$.
Here, $v_{\rho }$, $v_{z}$ and $v_{\phi }$ are the components of
the c.m velocity in the cylindrical coordinate and $v$ the speed
rate. On the other hand, the lower limit of $\left\vert E_{\pm
}-E_{0}\right\vert $ is $\left\vert \sqrt{\left\vert \Omega
_{c}\right\vert ^{2}+\left\vert \Omega _{p}\right\vert
^{2}+(V/2)^{2}}-V\right\vert $. Then the condition (\ref{ab}) can
be rewritten as
\begin{equation}
\frac{g}{r}\sqrt{\frac{2E}{M}}<<\frac{1}{2}(\sqrt{4\xi \left( r+\left\vert
z\right\vert \right) +E^{2}}-E).  \label{ab1}
\end{equation}%
Here we have used the fact that the upper limit of $v$ is
$\sqrt{2E/M}$ and the upper limit of $V$ is the atomic energy $E$.

The adiabatic condition (\ref{ab1}) can be satisfied in some
realistic cases. For instance, we confider the cesium atoms trapped
around the origin
$r=0$. We assume $\xi =\pi \times 10^{10}$ Hzm$^{-1/2}$ and the atomic energy $%
E\sim 10^{-26}$ J. Then it is easy to see that when the optical
angular
momentum $g\sim 10^{1}$, the adiabatic condition is satisfied when $%
r>10^{-6}$ m. Therefore, if the scale of the atomic ensemble is
$r\sim 10^{-3}$ m, the adiabatic approximation holds in almost all
the region of the atomic motion. Another case discussed above is
that the cesium atoms are trapped around the point $\left(
0,0,z_{0}\right) $ by the harmonic potentials. We assume $z_{0}\sim
10^{-3}$ m, $g\sim 10^{4}$ and the
frequencies of the trap potentials are $\omega _{z}\sim 10^{6}$ Hz and $%
\omega \sim 10^{2}$ Hz. Then the adiabatic condition can be met
again if $\xi $ and $E$ have the same values as mentioned above.
Therefore, the induced
change of zero point energy $-mg/\left( 4Mz_{0}^{2}\right) $ is about $%
10^{1}$ Hz. The frequency shift $\widetilde{\omega }-\omega $
caused by the monopole potential will have the same order with
$\omega $.

Before concluding this paper we make a remark on the results that
we have obtained. As we have shown, when the certain conditions $\left\vert
\Omega _{p}\right\vert ^{2}=\xi \left( r + z\right) $, $\left\vert
\Omega _{c}\right\vert ^{2}=\xi \left( r - z\right) $ is
satisfied, the singularity of the monopole potential can be
eliminated by dividing the space into two caps. To see the
universality in choosing the profile of light beam we make the
following change: $\left\vert \Omega _{c}\right\vert ^{2}=\xi
\left[ \left( 2\eta -1\right) r-z\right] $ where $\eta >1$. In this
case, the induced gauge potential will be ${\bf A}_{0}=g\left(
1+\cos \theta \right) \hat{e}_{\phi }/(2\eta r\sin \theta )$. Thus
the singularity of ${\bf A}_{0}$ can NOT be eliminated with the
above method when $g/\eta $ is not an
integer. This is because in this case we will have $\psi _{0b}\left( {\bf r}%
\right) =\psi _{0a}\left( {\bf r}\right) e^{-ig\phi /\eta }$ and
if $g/\eta $ is not an integer, $e^{-ig\phi /\eta }$ is not a
single valued function. We notice that a similar situation appears
when the Dirac's quantization condition is not satisfied
\cite{yang2}. This concluding remark shows the possibility to test
the subtle effect of magnetic monopole with manipulation of ultra
cold atoms.

{\it This work is supported by the NSFC and the knowledge Innovation Program
(KIP) of the Chinese Academy of Sciences. It is also founded by the National
Fundamental Research Program of China with No. 001GB309310. One (CPS) of the
authors thanks Y.B. Dai \ for useful discussions.}


\begin{references}
\bibitem[a]{email} Electronic address: suncp@itp.ac.cn

\bibitem[b]{www} Internet www site: http:// www.itp.ac.cn/\symbol{126}suncp

\bibitem{dirac} P. A. M. Dirac, Proc. Roy. Soc. A {\bf 133}, 60 (1931).

\bibitem{gut} G. 't Hooft, Nucl. Phys. B {\bf 79}, 276 (1974); A. M. Polyakov, JETP
Lett. {\bf 20}, 194 (1974).

\bibitem{yang1} T. T. Wu and C. N. Yang, Phys. Rev. D {\bf 12}, 384
(1975).

\bibitem{BPF} M. V. Berry, Proc. R. Soc. London A {\bf 392}, 45 (1984); A.
Shapere and F. Wilczek (Ed.), {\it Geometric Phases in Physics} (World
Scientific, Singapore, 1989).

\bibitem{FZ} Z. Fang et.al., Science {\bf 302}, 92 (2003).

\bibitem{WZK} J. Moody, A. Shapere, F. Wilczek and A. Zee, Phys. Rev.
Lett. {\bf 56}, 893 (1986); C. A. Mead, Phys. Rev. Lett. {\bf 59},
161 (1987).

\bibitem{boa} M. Born, R. Oppenheimer, Ann. Physik {\bf 84}, 457 (1930).

\bibitem{prd} C. P. Sun, M. L. Ge, Phys. Rev. D {\bf 41}, 1349 (1990).

\bibitem{EIT} S. E. Harris, Physics Today {\bf 50}, 36 (1997); M. D. Lukin,
Rev. Mod. Phys. {\bf 75}, 457 (2003).

\bibitem{Sun-prl} C. P. Sun, Y. Li, and X. F. Liu, Phys. Rev. Lett. {\bf 91}%
, 147903 (2003).

\bibitem{al} L. Allen, M. Padgett, and M. Babiker, Prog. Opt. {\bf 39}, 291
(1999); L. Allen, S. M. Barnett, and M. J. Padgett, {\it Optical
Angular Momentum} (Institute of Physics, Bristol, 2003).

\bibitem{fermion} G. Juzeliunas and P. Ohberg, cond-mat/0402317.

\bibitem{yang2} T. T. Wu and C. N. Yang, Nucl. Phys. B {\bf 107}, 365 (1976).

\bibitem{v=0} I. Tamm, Z. Phys. {\bf 71}, 141 (1931).
\end{references}
\end{document}